\documentclass[jmp,aps,showpacs,reprint]{revtex4-1}
\UseRawInputEncoding
\usepackage{graphicx}
\usepackage{dcolumn}
\usepackage{amsmath}
\usepackage{epsfig}
\RequirePackage{xspace}

\usepackage{relsize}

\begin{document}

\title{
\large \bfseries \boldmath Renormalization-Group Evolution of the None-local Matrix Element of $B$-meson In QCD and $B\to \pi$ transition form factor}

\author{Sheng L\"{u}} 
\author{Mao-Zhi Yang}\email{ Corresponding author: yangmz@nankai.edu.cn}
\affiliation{School of Physics, Nankai University, Tianjin 300071,P.R. China}

\date{\today}


\begin{abstract}
We consider the renormalization-group evolution of the matrix element of $\langle 0| \bar{q}(z)_\beta [z,0]b(0)_\alpha| \bar{B}\rangle$, which can be used to define the distribution amplitudes for $B$ meson and widely applied in studies of $B$ meson decays. The contribution to the renormalization constant of the non-local operator $\bar{q}(z)_\beta [z,0]b(0)_\alpha$ is considered up to one-loop order in QCD. Since the quark fields in this operator are not directly coupled fields, momentum can not flow freely through this non-local operator. Momentum involved in this operator can be treated stringently in coordinate space. We find that the ultraviolet divergences regulated by dimensional parameter $\epsilon$ cancel with each other, and the evolution effect vanishes. The matrix element $\langle 0| \bar{q}(z)_\beta [z,0]b(0)_\alpha| \bar{B}\rangle$ escapes from the renormalization-group evolution. We then apply the matrix element in calculating $B\to\pi$ transition form factor, where the matrix element is obtained by using the $B$ meson wave function calculated in QCD-inspired potential model. By comparing with experimental data for the semileptonic decay of $B\to \pi \ell\nu$ and light-cone sum rule calculation, we analyse the perturbative and non-perturbative contributions to $B\to\pi$ transition form factor in the frame work of perturbative QCD approach. We find that the effectiveness of the suppression of Sudakov factor to soft contribution depends on the end-point behavior of $B$ meson wave function, and with the $B$-meson wave function used in this work, soft contribution can not be well suppressed. The hard contribution to the $B\to\pi$ transition form factor is about 59\%, and soft contribution can be as large as 41\% in the naive calculation. To make the perturbative calculation reliable, a soft momentum cutoff in the calculation and soft form factor have to be introduced.

\end{abstract}

\maketitle
\section{Introduction}
Studies of $B$ meson decays are important for researching the source of $CP$ asymmetry and testing
the methods of treating the decay amplitude developed on the basis of QCD. Several methods, QCD factorization
\cite{QCDf1,QCDf2,QCDf3,QCDf4}, and perturbative QCD (PQCD) approach based on $k_T$-factorication \cite{PQCD1,PQCD2,PQCD3,PQCD4,PQCD5,PQCD6,PQCD7}, have been developed in the last two or three decades. Theoretically the $B$-meson distribution amplitudes defined through the matrix element
of the non-local operator between the $B$-meson and vacuum states, $\langle 0| \bar{q}(z)_\beta [z,0]b(0)_\alpha| \bar{B}\rangle$, are important quantities in the calculation of the decay amplitude of $B$ meson, where $[z,0]$ is the
Wilson line connecting the space-time coordinates of $b$ quark and the light anti-quark $\bar{q}$, $\alpha$ and
$\beta$ are the spinor indices of the quark fields. In principle, the matrix element $\langle 0| \bar{q}(z)_\beta [z,0]b(0)_\alpha |\bar{B}\rangle$ may depend on the energy scale $\mu$ if one-loop and/or higher order QCD corrections to the matrix element are taken into account. The evolution of the matrix element should accordingly lead to the scale-dependence of the distribution amplitudes of $B$ meson.

In this work we study the evolution of the matrix element $\langle 0| \bar{q}(z)_\beta [z,0]b(0)_\alpha \bar{B}\rangle$ by considering QCD correction up to one-loop order. There have been several works on the evolution of the
matrix element or the distribution amplitudes of $B$ meson existing in the literature \cite{PW1,PW2,GN,LN}, where momentum can flow freely through the effective operator $\bar{q}(z)[z,0]b(0)$ in the Feynman rule for the quark field operator \cite{GN,LN}. Here our work differs from these works in the literature by several points: 1) We treat the operator $\bar{q}(z) [z,0]b(0)$ completely as an non-local operator and perform the calculation in coordinate space, so that the non-local property of the operator can be treated stringently. 2) The operator $\bar{q}(z) [z,0]b(0)$ is only composite operator formed by two uncoupled quark fields, so momentum cannot transit through it freely. We treat it stringently in the coordinate space. So momentum flow can be treated properly, and the diagrams of one-loop level contribute differently compared with previous works in the literatures. Finally we apply the $B$ meson wave function obtained by solving the bound-state equation in QCD-inspired potential model into the calculation of the $B\pi$ transition form factors for the semileptonic decay $B\to \pi \ell\nu$ in PQCD approach. We find that, with the new wave function for $B$ meson, Sudakov suppression effect to soft contribution in the form factors becomes worse. Soft contribution to the form factor can not be negligible. A cutoff for the momentum flowing through the gluon and soft form factor has to be introduced. By comparing with experimental data and the calculation of the light-cone sum rule (LCSR) method, soft form factor can be obtained, which is about 23\% in the final result of the form factor.

The remainder of the paper is organised as follows: We calculate the renormalization constant of the composite non-local operator of two quarks in Sec. II. The renormalization-group equation of the matrix element $\langle 0| \bar{q}(z)_\beta [z,0]b(0)_\alpha|\bar{B}\rangle$ is derived and solved in Sec. III. Section IV is for the application of the $B$ meson wave function in the calculation of $B\to\pi$ transition form factor in PQCD approach, and the hard and soft contribution to the form factor are analysed by confronting our theoretical calculation with experimental data and LCSR result. Section V is a brief summary.

\section{The Renormalization Constant of the Composite Non-local Operator of Two Quarks}
The composite non-local operator of two-quark fields can be defined as
\begin{equation}
O(z)=\bar{q}_{\beta}(z)[z,0]b_{\alpha}(0)
\label{O1}
\end{equation}
where $\bar{q}(z)$ is the light anti-quark field at the space-time point $z$, $b(0)$ the heavy quark field
at the origin, $\alpha$ and $\beta$ the spinor indices of fermion fields, and $[z,0]$ the path-ordered exponential
\begin{eqnarray}\label{Wil-line}
[z,0]&=&{\cal P}exp[-ig_sT^a\int_0^z dz^\mu A_\mu^a(z)]\nonumber\\
     &=&{\cal P}exp[-ig_sT^a\int_0^1 d\alpha z^\mu A_\mu^a(\alpha z)]
\end{eqnarray}
where $g_s$ is the strong coupling constant, and $\alpha$ is the integral parameter from 0 to 1 here, which shall be continually used in the following without confusing with the spinor index. Summation with respect to color indices is understood in Eq. (\ref{O1}). Note that the two-quark fields in Eq. (\ref{O1}) are only two independent fields, which are aligned together to form the composite operator, and they are not coupled fields as usual in field theory. So no momentum can flow freely from one quark to the other.

\begin{center}
\begin{figure}[h]
\epsfig{file=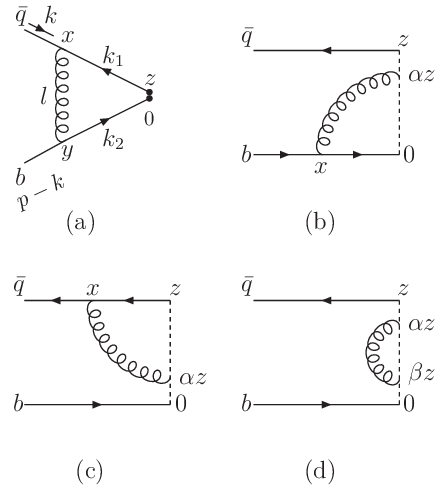,width=7cm,height=7cm} \caption{One-loop diagrams contributing to the matrix element
$\langle 0|O(\tilde{k})| \bar{q}^i(k)b^i(p-k)\rangle$. The vertex in (a) stands for the operator insertion of
$\bar{q}_\beta (z) b(0)_\alpha$, and the dashed lines in (b), (c) and (d) are the Wilson lines.} \label{fig1}
\end{figure}
\end{center}

The Fourier transform of the operator $O(z) $ is defined as
\begin{equation}
O(\tilde{k})=\int d^4 z e^{i\tilde{k}\cdot z}O(z)
\end{equation}
Then the matrix element of the operator between the vacuum state and $B$ meson in momentum space is
\begin{equation}
\langle O(\tilde{k})\rangle =\langle 0|O(\tilde{k}) | \bar{B}\rangle
\end{equation}
To derive the evolution of the matrix element, one needs to calculate the renormalization constant and anomalous
dimension of the operator $O(z)$. Therefore one should consider the matrix element of $O(z)$ sandwiched between the vacuum and free quark state of $\bar{q}^i(k)b^i(p-k)$ up to one-loop order in QCD. Here $i$ is the color index and the repeated indices mean summation over all the possible color states. To calculate the renormalization constant of the composite operator, one needs only to calculate the ultraviolet divergent terms. The diagrams contributing to the matrix element $\langle 0|O(\tilde{k})| \bar{q}^i(k)b^i(p-k)\rangle$ at one-loop order are shown in Fig.\ref{fig1}.

In the calculation of the matrix element of $\langle 0|O(\tilde{k})| \bar{q}^i(k)b^i(p-k)\rangle$, the path-ordered exponential factor given in Eq. (\ref{Wil-line}) needs also to be expanded perturbatively
\begin{eqnarray}\label{Wil-line2}
[z,0]&=&1-ig_sT^a\int_0^1 d\alpha z^\mu A_\mu^a(\alpha z)+\frac{1}{2}(-ig)^2T^aT^b\nonumber\\
     && \cdot\int_0^1d\alpha\int_0^1d\beta z^\mu z^\nu A_\mu ^a(\alpha z)A_\nu ^b (\beta z)    +\cdots
\end{eqnarray}
For the diagram of Fig.\ref{fig1}(a), only the leading term in the above expansion is needed. Since originally the operator $O(z)$ is a non-local composite operator composed by two quarks, it is convenient to calculate the contribution of the diagrams in Fig.\ref{fig1} in coordinate space. The matrix element can be generally written as
\begin{eqnarray} \label{Eexpansion}
\langle O(\tilde{k})\rangle &=&\langle 0|O(\tilde{k})|\bar{q}^i(k)b^i(p-k)\rangle \nonumber\\
&=&\int d^4 z e^{i\tilde{k}\cdot z}\langle 0 |O(z) |\bar{q}^i(k)b^i(p-k)\rangle\nonumber\\
&=&\int d^4 z e^{i\tilde{k}\cdot z}\langle 0 |\bar{q}_\beta(z)[1-ig_sT^a\int_0^1d\alpha \nonumber\\
&&    \cdot z^\alpha A_\mu^a(\alpha z)+\cdots]b_\alpha (0)|\bar{q}^i(k)b^i(p-k)\rangle
\end{eqnarray}
The contribution of Fig.\ref{fig1} (a) corresponds to the first term in the square bracket in Eq. (\ref{Eexpansion}).
With the correction of QCD in $\alpha_s$-order being considered, the contribution of Fig.\ref{fig1} (a) is
\begin{eqnarray}
T^{(1)}_a&=&\langle 0 |\mbox{T}\{\int d^4 z e^{i\tilde{k}\cdot z}
\int d^4 x ig_s (\bar{q}(x)\gamma_\mu A^{a\mu} (x)T^a q(x))\nonumber\\
&&\cdot\bar{q}(z)_\beta b(0)_\alpha ig_s\int d^4 y(\bar{q}(y)\gamma_\nu A^{a\nu} (y)T^a q(y))\}\nonumber\\
&&|\bar{q}^i(k)b^i(p-k)\rangle
\end{eqnarray}
Contracting the creation and annihilation operators in the quark state and field operators, one can finally
obtain the result of Fig.\ref{fig1}(a) as
\begin{eqnarray}
&&T^{(1)} _a= ig_s^2\frac{N_c^2-1}{2}\left[\bar{\upsilon}(k)\gamma_\mu
\frac{-\tilde{\not{k}}}{\tilde{k}^2+i\varepsilon}\right]_\beta \nonumber\\
&&\cdot\frac{1}{(\tilde{k}-k)^2+i\varepsilon}\left[\frac{\not{p}-\tilde{\not{k}}+m_b}{(p-\tilde{k})^2-m_b^2+i\varepsilon}
 \gamma^\mu u(p-k)\right]_\alpha\nonumber\\
&&
\end{eqnarray}
where $N_c$ is the number of color. This result shows that there is no ultraviolet divergence in the contribution of Fig.\ref{fig1}(a).

The contribution of Fig.\ref{fig1}(b) is
\begin{eqnarray}
T^{(1)}_b &=&\langle 0 |\mbox{T}\{\int d^D z e^{i\tilde{k}\cdot z}
\bar{q}(z)_\beta (-ig_s)T^a\int_0^1 d\alpha z^\mu A_\mu^a(\alpha z)\nonumber\\
&&\cdot b(0)_\alpha\int d^4 x ig_s (\bar{b}(x)\gamma_\nu T^b b(x)) A^{b\nu} (x)\}\nonumber\\
&&|\bar{q}^i(k)b^i(p-k)\rangle
\end{eqnarray}
Contracting the creation and annihilation operators in the field operators and that in the two-quark state,
we can obtain
\begin{eqnarray}
&&T^{(1)}_b =\int d^D z e^{i\tilde{k}\cdot z}\bar{\upsilon}_\beta (k)e^{-ik\cdot z}
(-ig_s)T^a\int_0^1d\alpha z^\mu \nonumber\\
&&\cdot\int\frac{d^D l}{(2\pi)^D}\frac{i}{l^2+i\varepsilon}e^{-i(\alpha z-x)\cdot l}
ig_s\left[\int d^Dx\int\frac{d^D k_b}{(2\pi)^D}\right.\nonumber\\
&&\left.\cdot\frac{i(\not{k}_b+m_b)}{k_b^2-m_b^2+i\varepsilon}e^{ix\cdot k_b}\gamma_\mu u(p-k)e^{-i(p-k)\cdot x}T^a\right]_\alpha
\end{eqnarray}
The space-time coordinate $z^\mu$ in the above equation is treated as
\begin{equation} \label{z-trick}
    z^\mu e^{i\tilde{k}\cdot z}=\frac{\partial}{i\partial \tilde{k}_\mu} e^{i\tilde{k}\cdot z}
\end{equation}
After the integration with respect to $d^D z$, a delta function
is obtained
\begin{equation}
\delta^D(\tilde{k}-k-\alpha l)=\frac{1}{\alpha^D}\delta^D(\frac{\tilde{k}-k}{\alpha}-l)
\end{equation}
which indicates that $\alpha\to 0$ corresponds to the ultraviolet limit. The final result for the contribution
of Fig.\ref{fig1}(b) is calculated to be
\begin{eqnarray}
T^{(1)} _b&=&g_s^2\frac{N_c^2-1}{2}\frac{\pi^2}{\epsilon}\delta^4(\tilde{k}-k)
 \bar{\upsilon}_\beta (k) u_\alpha (p-k)\nonumber\\
 &&+\mbox{finite terms}
\end{eqnarray}
where $\epsilon$ is the dimensional parameter defined by $D=4-2\epsilon$, and to get the above result,
the following relation is used
\begin{equation}
\lim_{\varepsilon\to 0}\frac{\varepsilon}{[(\tilde{k}-k)^2+i\varepsilon]^3}=\frac{\pi^2}{2}\delta^4(\tilde{k}-k)
\end{equation}

The calculation of the diagram of Fig.\ref{fig1}(c) is similar to that of Fig.\ref{fig1}(b), but the ultraviolet
limit occurs as $\alpha\to 1$ for this diagram. The result for the ultraviolet divergence is the same as that of Fig.\ref{fig1}(b).

The calculation of Fig.\ref{fig1}(d) involves the contribution of the second-order expansion of $[z,0]$
in Eq. (\ref{Wil-line2}). The contribution is
\begin{eqnarray}
&&T^{(1)} _d=\langle 0|\mbox{T}\{\int d^D z e^{i\tilde{k}\cdot z}\bar{q}_\beta(z)\frac{1}{2!}
  (-ig_s)^2 T^a T^b\nonumber\\
&&\cdot\int_0^1 d\alpha \int_0^1 d\beta z^\mu z^\nu A_\mu^a(\alpha z)A_\nu^b(\beta z)b_\alpha (0)\}
  |\bar{q}^i(k)b^i(p-k)\rangle\nonumber\\
\end{eqnarray}
Contracting the relevant creation and annihilation operators, we have
\begin{eqnarray}
&&T^{(1)} _d=-\frac{1}{2}g_s^2\frac{N_c^2-1}{2}\int d^Dz \int_0^1d\alpha\int_0^1d\beta
 \bar{\upsilon}_\beta (k)\nonumber\\
 &&~~\cdot z^2 e^{i\tilde{k}\cdot z}e^{-ik\cdot z}\int\frac{d^D l}{(2\pi)^D}
 \frac{i}{l^2+i\varepsilon}e^{-i(\alpha-\beta)z\cdot l} u_\alpha(p-k) \nonumber\\
 &&~~\label{eab0}
\end{eqnarray}
Using the trick in Eq. (\ref{z-trick}), the above equation can be changed to be
\begin{eqnarray}
&&T^{(1)} _d=\frac{i}{2}g_s^2\frac{N_c^2-1}{2}\int_0^1d\alpha\int_0^1d\beta\int d^D l
\frac{1}{(\alpha-\beta)^D}\nonumber\\
 &&~~\cdot \frac{\partial^2}{\partial \tilde{k}^\mu \partial \tilde{k}^\nu}
 \delta^D\left(\frac{\tilde{k}-k}{\alpha-\beta}-l\right)
 \frac{1}{l^2+i\varepsilon}\bar{\upsilon}_\beta (k)u_\alpha(p-k) \nonumber\\
 &&~~\label{eab1}
\end{eqnarray}
The $\delta$-function in Eq. (\ref{eab1}) indicates that $\alpha -\beta \to 0$ gives the ultraviolet limit, which
contributes the ultraviolet divergence. After integrating over momentum $l$, one gets
\begin{eqnarray}
&&T^{(1)} _d=\frac{i}{2}g_s^2\frac{N_c^2-1}{2}\int_0^1d\alpha\int_0^1d\beta
\frac{1}{(\alpha-\beta)^{D-2}}\nonumber\\
 &&~~\cdot \Bigg(\frac{2(4-D)}{[(\tilde{k}-k)^2+i\varepsilon]^2}-\frac{8i\varepsilon}{[(\tilde{k}-k)^2+i\varepsilon]^3}
 \bigg)    \bar{\upsilon}_\beta (k)u_\alpha(p-k) \nonumber\\
 &&~~\label{eab}
\end{eqnarray}
The ultraviolet divergence can be obtained by performing the integration over $\alpha$ and $\beta$. The result is
\begin{eqnarray}
T^{(1)} _d&=&g_s^2(N_c^2-1)(\bigg(\frac{2}{\delta}-\frac{1}{\epsilon}\bigg)\pi^2\delta^4(\tilde{k}-k)
 \bar{\upsilon}_\beta (k) u_\alpha (p-k)\nonumber\\
 &&+\mbox{finite terms}
\end{eqnarray}
where $\delta$ is an infinitesimal parameter appearing in the integration of $\alpha$ and $\beta$. $\delta$ should be finally taken to be $\delta\to 0$ after the limit $\epsilon\to 0$ is taken. Therefore $\delta$ is an extra regulator for the ultraviolet divergence other than the dimensional parameter $\epsilon$ in $D=4-2\epsilon$. The details of the integration about $\alpha$ and $\beta$ are given in Appendix \ref{a}.

The total contribution including the ultraviolet divergence at one-loop level contributed by the diagrams in Fig.\ref{fig1} is
\begin{eqnarray} \label{one-loop-u}
T^{(1)} &=&g_s^2(N_c^2-1)\pi^2\frac{2}{\delta}\delta^4(\tilde{k}-k)
 \bar{\upsilon}_\beta (k) u_\alpha (p-k)\nonumber\\
 &&+\mbox{finite terms}
\end{eqnarray}

To consider the renormalized matrix element, the renormalization constants $z_b^{1/2}$ and $z_q^{1/2}$
for the heavy quark $b$ and light quark $q$ must be included, which take care of the renormaliztion of the quark
fields. Then the relation between the bare and renormalized matrix element up to one-loop order  is
\begin{equation} \label{brmatrix}
\langle O(\tilde{k})\rangle^{(b)} =z_b^{-1/2} z_q^{-1/2}\frac{1}{(2\pi)^4}\int d^4k' Z_R(\tilde{k},k')\langle O_R(k')\rangle^{(r)}
\end{equation}
where $Z_R(\tilde{k},k')$ is the renormalization constant of the operator $O(\tilde{k})$ defined as
\begin{equation} \label{RC}
O(\tilde{k})=\frac{1}{(2\pi)^4} \int d^4k'Z_R(\tilde{k},k')O_R(k')
\end{equation}
Considering
\begin{equation}
z_b=z_q=1-\frac{\alpha_s}{4\pi}C_F\frac{1}{\epsilon}
\end{equation}
where $C_F=\frac{N^2_c-1}{2N_c}$, and the unrenormalized matrix element $\langle O(\tilde{k})\rangle^{(b)}$ up to one-loop
order
\begin{equation}
 \langle O(\tilde{k})\rangle^{(b)}=T^{(0)}+T^{(1)}
 \end{equation}
where $T^{(0)}$ is the tree-level matrix element
\begin{eqnarray}
T^{(0)}&=&\langle O(\tilde{k})\rangle^{(0)}\nonumber\\
 &=&\int d^4z e^{i\tilde{k}\cdot z}\langle 0|\bar{q}_\beta (z)b_\alpha (0)|\bar{q}^i(k)b^i(p-k)\rangle^{(0)} \nonumber\\
 &=& N_c(2\pi)^4\delta^4(\tilde{k}-k) \bar{\upsilon}_\beta (k) u_\alpha (p-k)
\end{eqnarray}
and $T^{(1)}$ is given in Eq. (\ref{one-loop-u}),  we can obtain the renormalization constant
\begin{equation} \label{ZR}
Z_R(\tilde{k},k)=\left[1+\frac{\alpha_s}{4\pi}C_F\bigg(\frac{4}{\delta}-\frac{1}{\epsilon}\bigg)\right]
(2\pi)^4\delta^4(\tilde{k}-k)
\end{equation}

\section{The Renormalization Group Equation and Its Solution}

The unrenormalized matrix element of the bare operator $\langle 0|O(\tilde{k})|\bar{B}\rangle^{(b)} $ should be scale-independent, so that
\begin{equation}
\mu\frac{d}{d\mu}\langle 0|O(\tilde{k})|\bar{B}\rangle^{(b)} =0
\end{equation}
Considering the relation between the bare and renormalized matrix element given in Eq. (\ref{brmatrix}), one can
get the renormalization-group equation for the matrix element $\langle 0|O_R(\tilde{k})|\bar{B}\rangle^{(r)} $
\begin{eqnarray}
&&\mu\frac{d}{d\mu}\tilde{\Phi}_{\alpha\beta}(k'',k,\mu)
+\int d^4k'\gamma(k'',k',\mu)\tilde{\Phi}_{\alpha\beta}(k',k,\mu)\nonumber\\
&& -\gamma_F \tilde{\Phi}_{\alpha\beta}(k'',k,\mu)=0
\end{eqnarray}
where
\begin{equation}
\tilde{\Phi}_{\alpha\beta}(k'',k,\mu)=\langle 0|O_R(k'')|\bar{B}(k)\rangle^{(r)}
\end{equation}
and $\gamma_F=\frac{1}{z_F} \mu\frac{d}{d\mu}z_F=2C_F \frac{\alpha_s}{4\pi}$ is the anomalous dimension of quark fields. $\gamma(k'',k',\mu)$ is the anomalous dimension of the operator $O(k'')$, which is defined as
\begin{equation}
\gamma(k'',k',\mu)=\int d^4\tilde{k} Z_R^{-1}(k'',\tilde{k},\mu)\mu \frac{d}{d\mu}Z_R(\tilde{k},k',\mu)
\end{equation}
and $Z_R^{-1}(k'',\tilde{k},\mu)$ is defined by
\begin{equation}
\int d^4\tilde{k}Z_R^{-1}(k'',\tilde{k},\mu)Z_R(\tilde{k},k',\mu)=\delta^4(k''-k')
\end{equation}
With the renormalization constant obtained in Eq. (\ref{ZR}), we can get the anomalous dimension
\begin{equation}
\gamma(k'',k',\mu)=2C_F \frac{\alpha_s}{4\pi}\delta^4(k''-k')
\end{equation}
With the above result, the renormalization-group equation is reduced to
\begin{equation}
\mu\frac{d}{d\mu}\tilde{\Phi}_{\alpha\beta}(k'',k,\mu)=0
\end{equation}
which indicates the matrix element for the $B$ meson is scale-independent. This result is different from
that obtained in Refs. \cite{GN,LN}. The reasons causing the difference are two-folds. One is that momentum can not
flow freely through the composite operator $\bar{q}(z)_\beta [z,0]b(0)_\alpha$ in our treatment, because the quark
fields in this non-local composite operator are not coupled fields as usual. The other is that we did not take the light-cone approximation, which is relevant to the contribution of Fig.\ref{fig1} (d). We find that Fig.\ref{fig1} (d) does contribute  ultraviolet divergence, and the term regulated by the dimensional parameter exactly cancels that of the other diagrams in Fig.\ref{fig1}. The renormalization behavior of the matrix element for $B$ meson is also different from that of distribution amplitudes for light mesons, which is obtained from Brodsky-Lepage kernel \cite{BL}, and where the light-meson distribution amplitude is defined by integrating the transverse momentum up to a specific scale $\mu$.

The matrix element $\langle 0| \bar{q}(z)_\beta [z,0]b(0)_\alpha \bar{B}\rangle$ plays an important
role in studying $B$-meson decays. It is in principle of non-perturbative dynamics. In the following, we shall consider the $B$-meson wave function obtained in the QCD-inspired relativistic potential model \cite{Yang2012,LY2014,LY2015}, a model inspired by the properties of QCD. Then the matrix element was derived in Ref. \cite{SY2017}. It is interesting to apply this matrix element in $B$ decays. The matrix element given in Ref. \cite{SY2017} is
\begin{eqnarray}
  \Phi_{\alpha\beta}(&k&)=\frac{-if_Bm_B}{4}K(\vec{k})
\nonumber\\
 && \cdot\Bigg\{(E_Q+m_Q)\frac{1+\not{v}}{2}\Bigg[\Bigg(\frac{k^+}{\sqrt{2}}  +\frac{m_q}{2}\Bigg)\not{n}_+
\nonumber\\
&&+\Bigg(\frac{k^-}{\sqrt{2}}  +\frac{m_q}{2}\Bigg)\not{n}_- -k_{\perp}^{\mu}\gamma_{\mu}  \Bigg]\gamma^5\nonumber\\
&&-(E_q+m_q)\frac{1-\not{v}}{2} \Bigg[  \Bigg(\frac{k^+}{\sqrt{2}}-\frac{m_q}{2}\Bigg)\not{n}_+
\nonumber\\
 && +\Bigg(\frac{k^-}{\sqrt{2}}-\frac{m_q}{2}\Bigg)\not{n}_--k_{\perp}^{\mu}\gamma_{\mu}\Bigg]\gamma^5
  \Bigg\}_{\alpha\beta}\label{eqm}
\end{eqnarray}
where $m_Q$ is the mass of the heavy quark, $m_Q=m_b$, $m_q$ the mass of the light quark, $k$ the momentum of the
light quark in the rest frame of the meson, $E_Q$ and $E_q$ the energies of the heavy and light quarks respectively,
$v$ the four-speed of the $B$ meson which satisfies $p_B^\mu=m_B v^\mu$, $n_\pm^\mu$ are two light-like vectors $n_\pm^\mu=(1,0,0,\mp 1)$, and
\begin{equation}
k^\pm=\frac{E_q\pm k^3}{\sqrt{2}},\;\;\; k_\perp^\mu=(0,k^1,k^2,0) \label{kpm}
\end{equation}
$K(\vec{k})$ is a function proportional to the $B$-meson wave function
\begin{equation}
K(\vec{k})=\frac{2N_B\Psi_0(\vec{k})}{\sqrt{E_qE_Q(E_q+m_q)(E_Q+m_Q)}} \label{wave-k}
\end{equation}
where $N_B$ is a factor defined as $N_B=\frac{i}{f_B}\sqrt{\frac{3}{(2\pi)^3m_B}}$, and $\Psi_0(\vec{k})$ is the wave function of $B$ meson, which is normalized as $\int d^3k |\Psi_0(\vec{k})|=1$. $\vec{k}$ is the 3-momentum of the light quark in the rest frame of the meson. The wave function $\Psi_0(\vec{k})$ has been solved numerically in Refs. \cite{Yang2012,LY2014,LY2015}, and an analytical form is given in Refs. \cite{SY2017,SY2019} by fitting to the numerical solution in the QCD-inspired potential model
\begin{equation}\label{psi0}
\Psi_0(\vec{k})=a_1 e^{a_2|\vec{k}|^2+a_3|\vec{k}|+a_4}
\end{equation}
where $a_i$ ($i=1,\cdots, 4$) are the parameters obtained as \cite{SY2019}
\begin{eqnarray}
&&a_1=1.66_{-0.11}^{+0.07}\,\mathrm{GeV}^{-3/2},\quad\;
  a_2=-1.07_{-0.16}^{+0.12}\,\mathrm{GeV}^{-2},\nonumber\\
 && a_3=-0.98_{-0.12}^{+0.17}\,\mathrm{GeV}^{-1},\quad   a_4=-0.130_{-0.04}^{+0.02}
\end{eqnarray}
for $B$-meson, and
\begin{eqnarray}
 &&a_1=1.97_{-0.09}^{+0.05}\,\mathrm{GeV}^{-3/2},\quad
  a_2=-1.09_{-0.18}^{+0.12}\,\mathrm{GeV}^{-2};\nonumber\\
 &&a_3=-0.69_{-0.10}^{+0.07}\,\mathrm{GeV}^{-1},\quad
  a_4=-0.47_{-0.07}^{+0.03}
\end{eqnarray}
for $B_s$ meson.

It deserves to note that the dependence on the transverse and longitudinal momenta $\vec{k}_\perp =(k^1,k^2)$ and $k^3$ is given simultaneously in Eqs. (\ref{wave-k}) and (\ref{psi0}), which is the solution of the bound-state equation in the QCD-inspired potential model.


\section{The application of the $B$-meson matrix element to the Transition Form Factor of $B\to \pi \ell \nu$ decay}
In this section the $B$-meson wave function obtained in QCD-inspired relativistic potential model will be used to calculate the $B\to \pi$ transition form factor in the frame work of PQCD \cite{PQCD1,PQCD2,PQCD3,PQCD4,PQCD5,PQCD6,PQCD7}, where the transverse momentum and Sudakov factor are included. The Sudakov factor can suppress the contribution of soft dynamics to ensure the applicability of PQCD approach. We shall study how large the soft contribution is left if the new wave function of $B$ meson is used.

The $B\to \pi$ transition form factors are defined by the hadronic matrix element
\begin{eqnarray}
&&\;\;\;\; \langle \pi (p_2)\mid V_\mu\mid \bar{B}(p_1)\rangle\nonumber\\
&&=F_+^{B\pi}(q^2)(p_1+p_2-\frac{m^2_{B}-m_\pi^2}{q^2}q)_\mu\nonumber\\
&&\;\;\;\;+F_0^{B\pi}(q^2)\frac{m^2_{B}-m_\pi^2}{q^2}q_\mu, \label{matrix-el}
\end{eqnarray}
where $q=p_1-p_2$, and $p_{1,2}$ are the momenta of $B$ and $\pi$ mesons, respectively. For $q^2=0$, there is $F_+^{B\pi}(0)=F_0^{B\pi}(q^2)$. For the semileptonic decay $B\to \pi \ell \nu$, where $\ell =e$ or $\mu$, the contribution of the form factor $F_0^{B\pi}(q^2)$ can be dropped for the sake of its being proportional to the small mass squared of the lepton $m_\ell^2$. The differential decay width of a semileptonic decay $B\to \pi\ell\nu$ can be calculated to be
\begin{equation}
\frac{d\Gamma}{dq^2}(B\to \pi\ell\nu)=\frac{G_F^2}{24\pi^3}|V_{ub}|^2|F_+^{B\pi}(q^2)|^2|\vec{p}_\pi|^3. \label{width-e}
\end{equation}
where $G_F$ is the Fermi constant, $V_{ub}$ the Cabibbo-Kobayashi-Maskawa (CKM) matrix element, and $\vec{p}_\pi$ the momentum of the pion in the rest frame of $B$ meson.

\begin{center}
\begin{figure}[h]
\epsfig{file=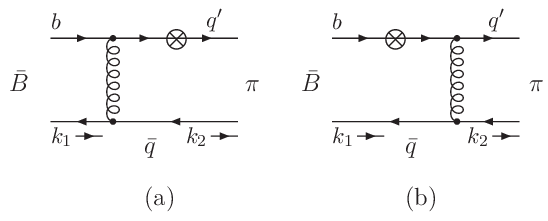,width=8cm,height=3cm} \caption{Diagrams contributing to $B\to \pi$ transition
form factors, where the cross-circle $\otimes$ denotes the weak vertex.} \label{fig2}
\end{figure}
\end{center}

In the large recoil region where $q^2\sim 0$, large momentum and energy are released in the decays of $B$ meson to light particle states. Therefore the form factors are mainly controlled by short-distance interactions in this region and PQCD can be applied in large recoil region \cite{PQCD1,liyu1996-1,luyang2003}. Factorization theorem plays an important role in applying perturbative QCD in hadronic $B$ decays, where soft interactions are separated from the hard interaction. The soft inactions are absorbed into hadronic wave functions, and hard interaction is calculated by perturbation theory \cite{liyu1996-1}. The Feynman diagrams for the hard interaction in $B\to \pi$ transition form factor at $\alpha_s$ order are depicted in Fig.\ref{fig2}.

A direct calculation of the diagrams in Fig.\ref{fig2} suffers from the infrared singularity as the gluon momentum tends to the soft limit, which is called the end-point singularity. To cure the end-point singularity, partons' transverse momentum $k_\perp$ is introduced, and double logarithmic contributions such as $\alpha_s \mbox{ln}^2 k_\perp$ in higher order corrections are summed to all orders by the technique of resummation, which gives the Sudakov factor \cite{liyu1996-1,liyu1996-2}. In addition, there are also double logarithms such as $\alpha_s \mbox{ln}^2 x$ in the higher order corrections, where $x$ is the parton's momentum fraction. These double logs should also be summed to all order, which gives the threshold resummation \cite{lihn2002}. Both the Sudakov and threshold resummation factors can suppress the long-distance contribution, so that the end-point singularity can be cured, which improves the applicability of perturbation calculation.

In Refs. \cite{Yang2012,LY2014,LY2015}, the $b$-flavored heavy-light quark-antiquark bound-states are studied in QCD-inspired relativistic potential model. The whole mass spectrum of $B$-system including higher excited states are obtained, which are consistent with experimental data within a few percent. Therefore, $B$ meson can be safely viewed as dominantly quark-antiquark bound-state, and contributions of the higher Fock states should be less than a few percent. In this work we only consider the leading Fock state and higher Fock states are neglected as a good approximation.

The factorization formula for the $B\pi$ transition matrix element has been previously proved and given in Ref. \cite{liyu1996-1}. The formula in coordinate space and with spinor indices explicitly given is \cite{luyang2003}
\begin{eqnarray}
&&\;\;\;\; \langle \pi (p_2)\mid V_\mu\mid \bar{B}(p_1)\rangle\nonumber\\
&&=\int d^4zd^4y \langle \pi(p_2)|\bar{q}'(0)_\rho q(y)_\delta |0\rangle\nonumber\\
&&\;\;\times H_\mu ^{\rho\alpha;\beta\delta}(y,z)\langle 0|\bar{q}_\beta (z) b_\alpha (0)|\bar{B}\rangle \label{fac}
\end{eqnarray}
where $H_\mu ^{\rho\alpha;\beta\delta}(y,z)$ stands for the hard transition amplitude. The hadronic matrix elements
$\langle 0|\bar{q}_\beta (z) b_\alpha (0)|\bar{B}\rangle$ and $\langle \pi(p_2)|\bar{q}'(0)_\rho q(y)_\delta |0\rangle$ define the hadronic distribution amplitudes. For the matrix element of $B$ meson, we have
\begin{equation}
\langle 0|\bar{q}_\beta (z) b_\alpha (0)|\bar{B}\rangle =\int d^3 k_1\Phi_{\alpha\beta}(k_1,\mu)e^{-ik_1\cdot z}
\end{equation}
where $\Phi_{\alpha\beta}(k_1)$ is the function given in Eq. (\ref{eqm}). Here, we will check the phenomenological effect of the $B$ meson wave function obtained by solving bound-state equation in our previous works, which works for describing the whole mass-spectrum of $b$-flavored system \cite{Yang2012,LY2014,LY2015}. We will check the applicability of perturbative QCD in the calculation of $B\pi$ transition form factor with the new wave function for $B$ meson being used.

For the matrix element of $\pi$ meson, which is fast moving in the large-recoil limit, the light-cone distribution amplitudes can be defined as \cite{bra1990,bal1999}
\begin{eqnarray}
\langle \pi(p_2)|\bar{q}'(0)_\rho q(y)_\delta |0\rangle &=&\int dx d^2k_{2\perp}e^{ix p_2\cdot y-y_{\perp}\cdot k_{2\perp}}\nonumber\\
&&\times\Phi^\pi_{\delta\rho}
\end{eqnarray}
with the spinor wave function $\Phi^\pi_{\delta\rho}$ to be
\begin{eqnarray}
\Phi^\pi_{\delta\rho}&=&\frac{if_\pi}{4}\Bigg\{\not{p}_2\gamma_5\phi_\pi(x,k_{2\perp})
  -\mu_\pi\gamma_5\Bigg(\phi^\pi_P(x,k_{2\perp})\nonumber\\
   && -\sigma_{\mu\nu}p_2^\mu y^\nu\frac{\phi^\pi_\sigma(x,k_{2\perp})}{6}\Bigg)\Bigg\}_{\delta\rho}
\end{eqnarray}
where $f_\pi$ is the decay constant of pion, $\phi_\pi$, $\phi^\pi_P$ and $\phi^\pi_\sigma$ are twist-2 and twist-3 distribution functions of pion, and $\mu_\pi=m^2_\pi/(m_u+m_d)$ for the charged pion. In the momentum space, $\Phi^\pi_{\delta\rho}$ can be changed to be \cite{bf2001,wy2002}
\begin{eqnarray}
\Phi^\pi_{\delta\rho}&=&\frac{if_\pi}{4}\Bigg\{\not{p}_2\gamma_5\phi_\pi(x_2,k_{2\perp})
  -\mu_\pi\gamma_5\Bigg(\phi^\pi_P(x_2,k_{2\perp})\nonumber\\
   && -i\sigma_{\mu\nu}\frac{p_2^\mu \bar{p}_2^\nu}{p_2\cdot \bar{p}_2} \frac{\phi'^\pi_\sigma(x_2,k_{2\perp})}{6}\nonumber\\
   &&+i  \sigma_{\mu\nu}p_2^\mu\frac{\phi^\pi_\sigma(x_2,k_{2\perp})}{6}\frac{\partial}{\partial k_{\perp\nu}}\Bigg)\Bigg\}_{\delta\rho}
\end{eqnarray}
where $\bar{p}_2$ is a 4-momentum with its vector direction opposite to the moving of pion, and $\phi'^{\pi}_\sigma (x,k_{2\perp})=\frac{\partial \phi^\pi_\sigma (x,k_{2\perp})}{\partial x}$.

In the calculation, the momenta are taken in the light-cone coordinate, i.e., a momentum $A^\mu$ is written as $A^\mu=(A^+,A^-,\vec{A}_\perp)$, with $A^\pm=\frac{A^0\pm A^3}{\sqrt{2}}$. The scalar product of two arbitrary momenta $A^\mu$ and $B^\mu$ can be calculated as $A\cdot B=A^\mu B_\mu=A^+B^-+A^-B^+-\vec{A}_\perp\cdot \vec{B}_\perp$. The momentum of $B$ meson in the rest-frame is $p_1^\mu=\frac{1}{\sqrt{2}}(m_B,m_B,\vec 0_\bot)$. The coordinate axes are taken in such a way that the $\pi$-meson moves in the opposite direction of $z$ axis, then the momentum of pion is $p_2^\mu=(0,\eta m_B/\sqrt{2},\vec 0_\bot)$, where the mass of pion is dropped, and $\eta$ is a parameter that is related to the momentum squared of the lepton pair as $q^2=(1-\eta)m^2_B$. In addition, $\bar{p}_2^\mu=(\eta m_B/\sqrt{2},0,\vec 0_\bot)$. The momentum of the light quark in $B$ meson can be written in terms of the light-cone component as
\begin{equation}
k_1=(x_1 \frac{m_B}{\sqrt{2}}, x'_1 \frac{m_B}{\sqrt{2}}, \vec{k}_{1\perp}) \label{x12}
\end{equation}
Comparing the above equation with Eq. (\ref{kpm}), one can get
\begin{equation}
x_1=\frac{E_q+k^3_1}{m_B},\;\;\; x_1'=\frac{E_q-k^3_1}{m_B}
\end{equation}
The momentum of the light quark in the $\pi$ meson is
\begin{equation}
k_2=(0,x_2 \eta \frac{m_B}{\sqrt{2}},\vec{k}_{2\perp})
\end{equation}
Both the fraction parameters $x_1$ and $x_1'$ are at the order of $\Lambda_{QCD}/m_B$, while $x_2$ is at the order of 0.5 in the large recoil limit. So $x_1$ and $x_1'$ are small parameters compared with $x_2$, and $x_1'$ can be dropped because it is always in substraction with $x_2$. So the hard amplitude only depends on $x_1$ and $x_2$, and $x_1'$ is neglected. PQCD approach works in this prescription.

 In Eqs. (\ref{eqm}) and (\ref{wave-k}), the wave function of $B$ meson is given as a function of the light-quark momentum in the rest frame of $B$ meson, which is spherically symmetric as a ground-state of quark-antiquark bound state. It can be transferred to be a function of $x_1$ and $\vec{k}_{1\perp}$ by using the definition of the fractional variable $x_1$, then the integral measure $d^3k_1$ can be transferred to
 \begin{equation}
 d^3k_1=(\frac{1}{2}m_B+\frac{|\vec{k}_{1\perp}|^2}{2x_1^2m_B})dx_1d^2 k_{1\perp }
 \end{equation}
 To derive this measure transformation, the mass of the light-quark is neglected. To keep the momentum conservation between the $B$ meson and its quark components, the light quark is treated to near its mass-shell, while the heavy quark can be far off-shell in the decay process, then the energy of the heavy quark can be obtained by $E_Q=m_B-E_q$. The off-shell mass of the heavy quark can be obtained as a function of $k_1$
 \begin{equation}
m_Q(k_1)=\sqrt{E_Q^2-|\vec{k}_1|^2}=\sqrt{m_B^2-2|\vec{k}_1|m_B}
 \end{equation}
 To keep the heavy-quark mass to be real, we restrict
 \begin{equation}
 |\vec{k}_1|\le\frac{1}{2}m_B
 \end{equation}
 Certainly, beyond this upper limit, the wave function effectively vanishes \cite{Yang2012,LY2014,LY2015}. From this upper limit, we can get the limits of $x_1$
 \begin{equation}
 x_1^d\le x_1\le x_1^u, \;\;\; x_1^{u,d}=\frac{1}{2}\pm \sqrt{\frac{1}{4}-\frac{|\vec{k}_{1\perp}|^2}{m_B}}
 \end{equation}

The next-to-leading order corrections to the hard transition amplitude in the $B\pi$ form factor was investigated in Ref. \cite{lsw2012} several years ago. Here we restrict ourself in the leading-order contribution in this work, because the power of soft contribution may change with the new wave function for $B$ meson being used. It is necessary to investigate the phenomenological effect of the new wave function at leading-order at first. The calculation of the transition matrix element of Eq. (\ref{fac}) is performed in $b$-space ($\vec{b}_\bot$ being Fourier transform variable of $\vec{k}_{\perp}$). The form factors can be extracted from this calculation. The result of $F_{+,0}^{B\pi}(0)$ is
\begin{eqnarray}
&&F_+^{B\pi}(0)=F_0^{B\pi}(0)\nonumber\\
&&=\frac{2\pi^2}{N_c^2}f_B f_\pi m_B\int dk_{1\perp}k_{1\perp}\int_{x_1^d}^{x_1^u}dx_1\int_0^1 dx_2
\int_0^\infty b_1db_1b_2db_2 \nonumber\\
&&\times (\frac{1}{2}m_B+\frac{|\vec{k}_{1\perp}|^2}{2x_1^2m_B})K(\vec{k}_1)(E_Q+m_Q)J_0(k_{1\perp}b_1)\nonumber\\
&&\Bigg\{\alpha_s(\mu_{e1})
\Bigg(2m_B[E_q(1+x_2)+k_1^3(1-x_2)]\phi_\pi(x_2,b_2)\nonumber\\
&&+2\mu_\pi[E_q(1-2x_2)-k_1^3]\phi^\pi_P(x_2,b_2)\nonumber\\
&&+\frac{1}{3}\mu_\pi[E_q(2 x_2-1)+k_1^3]\phi'^\pi_\sigma(x_2,b_2)-\frac{4}{3}\mu_\pi \phi^\pi_\sigma(x_2,b_2)]\Bigg)\nonumber\\
&&\times h_e(x_1,x_2,b_1,b_2)\exp[-S_{B}(\mu_{e1})-S_{\pi}(\mu_{e1})] \nonumber\\
&&+\alpha_s(\mu_{e2}) [4\mu_\pi(E_q-k_1^3)]\phi^\pi_P(x_2,b_2)\nonumber\\
&& \times h_e(x_2,x_1,b_2,b_1)\exp[-S_{B}(\mu_{e2})-S_{\pi}(\mu_{e2})]\Bigg\}
\label{eqform}
\end{eqnarray}
where the function $h_e$ is defined as
\begin{eqnarray}
&&h_e(x_1,x_2,b_1,b_2)=K_0(\sqrt{x_1 x_2}m_B b_1)\nonumber\\
&&\times [\theta (b_1-b_2)I_0(\sqrt{x_2}m_B b_2)K_0(\sqrt{x_2}m_B b_1)\nonumber\\
&&+\theta (b_2-b_1)I_0(\sqrt{x_2}m_B b_1)K_0(\sqrt{x_2}m_B b_2)]S_t(x_2)  \label{hardf}
\end{eqnarray}
with $J_0$ being the Bessel function of order 0, $I_0$ and $K_0$ the modified Bessel functions.

The functions $S_B$ and $S_\pi$ for the wave functions of $B$ meson and pion are given by \cite{PQCD3}
\begin{equation}
S_B(t) = s(x_1,b_1,m_B)-\frac{1}{\beta_1}\ln \frac{\ln (t/\Lambda_{\mbox{QCD}})}
         {\ln (1/(b_1\Lambda_{\mbox{QCD}}))}
\end{equation}
\begin{eqnarray}
S_\pi(t) &=& s(x_2,b_2,m_B)+s(1-x_2,b_2,m_B)\nonumber\\
&&\;\;-\frac{1}{\beta_1}\ln \frac{\ln (t/\Lambda_{\mbox{QCD}})}
         {\ln (1/(b_2\Lambda_{\mbox{QCD}}))}
\end{eqnarray}
The exponential $e^{-s(x,b,Q)}$ is the Sudakov factor, and the exponent $s(x,b,Q)$ has been given in Ref. \cite{PQCD3}. The explicit form of $s(x,b,Q)$ up to next-to-leading order is
\begin{widetext}
\begin{eqnarray}
&& s(x,b,Q)=\frac{A^{(1)}}{2\beta_{1}}\hat{q}\ln\left(\frac{\hat{q}}
{\hat{b}}\right)-
\frac{A^{(1)}}{2\beta_{1}}\left(\hat{q}-\hat{b}\right)+
\frac{A^{(2)}}{4\beta_{1}^{2}}\left(\frac{\hat{q}}{\hat{b}}-1\right)
-\left[\frac{A^{(2)}}{4\beta_{1}^{2}}-\frac{A^{(1)}}{4\beta_{1}}
\ln\left(\frac{e^{2\gamma_E-1}}{2}\right)\right]
\ln\left(\frac{\hat{q}}{\hat{b}}\right)
\nonumber \\
&&+\frac{A^{(1)}\beta_{2}}{4\beta_{1}^{3}}\hat{q}\left[
\frac{\ln(2\hat{q})+1}{\hat{q}}-\frac{\ln(2\hat{b})+1}{\hat{b}}\right]
+\frac{A^{(1)}\beta_{2}}{8\beta_{1}^{3}}\left[
\ln^{2}(2\hat{q})-\ln^{2}(2\hat{b})\right]
\nonumber \\
&&+\frac{A^{(1)}\beta_{2}}{8\beta_{1}^{3}}
\ln\left(\frac{e^{2\gamma_E-1}}{2}\right)\left[
\frac{\ln(2\hat{q})+1}{\hat{q}}-\frac{\ln(2\hat{b})+1}{\hat{b}}\right]
-\frac{A^{(2)}\beta_{2}}{16\beta_{1}^{4}}\left[
\frac{2\ln(2\hat{q})+3}{\hat{q}}-\frac{2\ln(2\hat{b})+3}{\hat{b}}\right]
\nonumber \\
& &-\frac{A^{(2)}\beta_{2}}{16\beta_{1}^{4}}
\frac{\hat{q}-\hat{b}}{\hat{b}^2}\left[2\ln(2\hat{b})+1\right]
+\frac{A^{(2)}\beta_{2}^2}{432\beta_{1}^{6}}
\frac{\hat{q}-\hat{b}}{\hat{b}^3}
\left[9\ln^2(2\hat{b})+6\ln(2\hat{b})+2\right]
\nonumber \\
&& +\frac{A^{(2)}\beta_{2}^2}{1728\beta_{1}^{6}}\left[
\frac{18\ln^2(2\hat{q})+30\ln(2\hat{q})+19}{\hat{q}^2}
-\frac{18\ln^2(2\hat{b})+30\ln(2\hat{b})+19}{\hat{b}^2}\right]
\label{sss}
\end{eqnarray}
\end{widetext}
where $\hat q$ and $\hat b$ are defined as
\begin{equation}
{\hat q} \equiv  {\rm ln}\left(xQ/(\sqrt 2\Lambda_{QCD})\right),~
{\hat b} \equiv  {\rm ln}(1/b\Lambda_{QCD})
\end{equation}
The coefficients $\beta_{i}$ and $A^{(i)}$ are
\begin{eqnarray}
& &\beta_{1}=\frac{33-2n_{f}}{12}\;,\;\;\;\beta_{2}=\frac{153-19n_{f}}{24}\; ,
A^{(1)}=\frac{4}{3}\;,
\nonumber \\
& & A^{(2)}=\frac{67}{9}-\frac{\pi^{2}}{3}-\frac{10}{27}n_
{f}+\frac{8}{3}\beta_{1}\ln\left(\frac{e^{\gamma_E}}{2}\right)\;
\end{eqnarray}
where $\gamma_E$ is Euler constant.

$S_t(x)$ in Eq. (\ref{hardf}) is the threshold resummation factor, which can be parameterized as \cite{lihn2002}
\begin{equation}
S_t(x)=\frac{2^{1+2c}\Gamma(3/2+c)}{\sqrt{\pi}\Gamma(1+c)} [x(1-x)]^c
\label{stx}
\end{equation}
where the parameter $c$ is determined to be $0.3$. The resummation factor vanishes efficiently fast at the end-point limit $x\to 0$ and $x\to 1$, so it helps to suppress the end-point contribution.

The wave functions of $\pi$ meson and the Fourier transform are given in the Appendix \ref{b}.

In the numerical analysis, with Eq. (\ref{eqform}), the form factor is calculated to be
$F_{+,0}^{B\pi}(0)=0.34\pm 0.01$,
where the uncertainty is estimated by varying the input parameters, including the parameters for $B$ meson wave function and  pion distribution amplitudes. The uncertainty caused by $B$ meson wave function is about 3.5\%, and 2.5\% caused by pion wave function. To check the effectiveness of the perturbative calculation, we can show the contribution varying with different values of $\alpha_s/\pi$, because QCD higher order correction usually occurs with the combination $\alpha_s/\pi$. The form factor varying with $\alpha_s/\pi$ is shown in Table \ref{table1}.

\begin{table}
\caption{\label{table1} Form factor $F_{+,0}^{B\pi}(0)$ with different values of $\alpha_s/\pi$. The last column is for the total contributions with all possible values of $\alpha_s/\pi$ without restriction. }
\begin{tabular}{|c|c|c|c|c|c|c|c|c|}
\hline\hline
  $\alpha_s/\pi$  & \small{$\le 0.1$} & \small{$\le 0.2$} & \small{$\le 0.3$} & \small{$\le 0.4$} & \small{$\le 0.5$} & \small{$\le 0.6$} & total \\ \hline
  \rule{0pt}{10pt}$F_{+,0}^{B\pi}(0)$ &0.06& 0.20& 0.26 & 0.28& 0.30 & 0.31 & 0.34 \\ \hline
  percentage& 18\% & 59\% & 75\% & 83\% & 87\% & 90\% & 100\%
 \\ \hline \hline
\end{tabular}
\end{table}

Table \ref{table1} shows that the contribution with $\alpha_s/\pi\le 0.2$ is 59\%, and $\alpha_s/\pi\le 0.3$ is 75\%. So the contribution with $\alpha_s/\pi>0.2$ and 0.3 can be as large as 41\% and 25\%, respectively. If we set a stringent perturbative criterion $\alpha_s/\pi< 0.2$, 41\% of the theoretical calculation will locate in the non-perturbative region, which makes the perturbaive calculation in the $k_T$ factorization fail. The reason is that we use the new $B$ wave function in the calculation in this work. We checked numerically that the end-point behavior of the $B$ wave function in Eq. (\ref{wave-k}) is $\sim \sqrt{x_1}$, which is less suppressive to the soft contribution at the end-point region $x_1\to 0$ than the $B$-meson wave function widely used in PQCD approach. The end-point behavior of the $B$-meson wave function widely used in PQCD approach is $\sim x_1^2$ as $x_1\to 0$. So the suppression to the soft contribution also relies on the meson wave function, other than Sudakov factor. Since the new wave function for $B$ meson has dynamical basis in deed, it is valuable to investigate the effect of it in $B$ meson decays. Note that the experimental data for the whole mass-spectrum of $b$-flavored meson system can be accommodated by the eigenvalues of the wave equation in the QCD-inspired potential model \cite{Yang2012,LY2014,LY2015}.

Now that the soft contribution to the form factor is sizable in our calculation, soft form factor has to be introduced as that given in the factorization approach proposed in Refs. \cite{bf2001,BF2004}
\begin{eqnarray}
F_{i}^{B\pi}&=&C_{i}\xi^{B\pi} +h_{i}^{B\pi} \\
h_{i}^{B\pi}&=&\Phi_B\otimes H \otimes \Phi_\pi
\end{eqnarray}
where $\xi^{B\pi}$ is the soft part of the form factor, $C_{i}=1+O(\alpha_s)$ and $i=+,0$. If the final value of $\xi^{B\pi}$ is small, $O(\alpha_s)$ can be neglected. $h_{i}^{B\pi}$ is the hard part of the form factor, and $\otimes$ stands for the convolution of the wave functions with the hard amplitude, which is relevant to Eq. (\ref{eqform}) with the scale being hard. By taking the hard criterion $\alpha_s/\pi<0.2$, we can get from Eq. (\ref{eqform})
\begin{equation}
h_{i}^{B\pi}=0.20\pm 0.01 \label{hard-form}
\end{equation}
The semileptonic decay $B\to \pi \ell\nu$ has been measured in experiments \cite{barbar2011,belle2011,barbar2012,belle2013}, and the $B\to\pi$ transition form factor has been investigated.
The combined fitted result including experimental data and LCSR result for the form factor at large recoil is \cite{belle2013,lcsr}
\begin{equation}
F_{+}^{B\pi}(0)=0.26\pm 0.02 \label{total-form}
\end{equation}
Comparing the above result of the form factor with the the hard contribution of the form factor given in Eq. (\ref{hard-form}), one can obtain the soft form factor
\begin{equation}
\xi^{B\pi}=0.06\pm 0.02 \label{soft-form}
\end{equation}
where the $O(\alpha_s)$ term in $C_i$ has been neglected in obtaining the above result, because the difference between the total and the hard part of form factor is small, which results in a small soft form factor. According to Eqs. (\ref{hard-form}), (\ref{total-form}) and (\ref{soft-form}), the hard dynamics contributes 77\% to the total form factor of $F_{+,0}^{B\pi}(0)$, while the soft part contributes 23\%. This result can help to understand the dynamical structure of the form factor, and give some hint for power counting for theoretical study of $B$ meson decays.

\section{Summary}
We study the renormalization-group evolution for the $B$-meson matrix element $\langle 0| \bar{q}(z)_\beta [z,0]b(0)_\alpha| \bar{B}\rangle$. The renormalization constant of the non-local operator $\bar{q}(z)_\beta [z,0]b(0)_\alpha$ is calculated up to one-loop order in QCD. Momentum can not flow through the non-local operator freely, because the quark fields in the operator are not directly coupled fields, which makes the result of the ultraviolet result simple. We did not make the light-cone approximation either, and we find that the ultraviolet divergences regularized by the dimensional parameter for different diagrams cancel with each other, which makes the evolution of the $B$-meson matrix element trivial. The renormalztion-group equation for the matrix element $\langle 0| \bar{q}(z)_\beta [z,0]b(0)_\alpha| \bar{B}\rangle$ is solved and the result is that the matrix element is scale-indpendent. The $B$ meson wave function obtained by solving bound-state equation and the relevant $B$-meson matrix element are applied in the calculation of $B\to\pi$ transition form factor for semileptonic decay $B\to \pi\ell\nu$ within the approach of $k_T$ factorization. The numerical calculation shows that the effectiveness of the suppression to the soft contribution also depends on the end-point behavior of the $B$-meson wave function other than Sudakov factor. In the case of the $B$-meson wave function obtained by solving the bound-state wave equation in QCD-inspired potential model, the suppression to the soft contribution is not effectively enough. Soft form factor has to be introduced. By comparing to the combined result of experimental data and light-cone sum rule calculation, we find the soft form factor is about 23\% of the total form factor, and the hard contribution is about 77\%.

\vspace{0.5cm}
\acknowledgments
This work is supported in part by the National Natural Science Foundation of China under
Contracts No. 11875168, 11375088.

\appendix{}
\section{\label{a}}
The integral about $\alpha$ and $\beta$ in Eq. (\ref{eab}) is
\begin{eqnarray}
I&\equiv &\frac{1}{2}\int_0^1 d\alpha \int_0^1 d\beta \frac{1}{(\alpha-\beta)^{D-2}}\nonumber\\
&=&\int_0^1 d\alpha \int_0^\alpha d\beta\frac{1}{(\alpha-\beta)^{D-2}}\nonumber\\
&=&\int_0^1 d\alpha \int_0^\alpha dx x^{2-D}
\end{eqnarray}
where $x=\alpha-\beta$. We do the integration in space of dimension $D=4-2\epsilon$. To make the integration regularized, it can be changed to
\begin{eqnarray}
I&=&\lim_{\delta\to 0} \int_0^1 d\alpha \int_\delta^\alpha dx x^{2-D}\nonumber\\
&=&\frac{1}{2\epsilon-1}\frac{1}{2\epsilon}-\frac{1}{2\epsilon-1}\delta^{2\epsilon -1}\nonumber \\
&=&-\frac{1}{2\epsilon}+\frac{1}{\delta}+O(\epsilon)+\cdots
\end{eqnarray}

\section{\label{b}}
The wave functions $\phi_\pi(x,k_{\perp})$, $\phi^\pi_P(x,k_{\perp})$ and $\phi^\pi_\sigma(x,k_{\perp})$ of $\pi$ meson are all assumed to be in a factorized form, where the transverse momentum dependence can be separated out, and assumed to a Gaussian distribution \cite{wy2002}
\begin{equation}
\phi_i(x,k_{\perp})=\phi_i(x)\times \Sigma(k_{\perp})
\end{equation}
with
\begin{equation}
 \Sigma(k_{\bot})=\frac{\beta^2}{\pi}\exp(-\beta^2 k_{\bot}^2)
\end{equation}
and $\phi_i(x,k_{\perp})$ stands for $\phi_\pi(x,k_{\perp})$, $\phi^\pi_P(x,k_{\perp})$ and $\phi^\pi_\sigma(x,k_{\perp})$ with $i=1,\; 2,\; 3$. Both $\phi_i(x)$ and $\Sigma(k_{\perp})$ are normalized to 1
\begin{equation}
\int_0^1\phi_i(x)=1,~~\int d^2 \vec k_{\bot}\Sigma(k_{\bot})=1
\end{equation}
When transform the wave functions into $b$-space, they become
\begin{eqnarray}  \label{eq:wavefunctionb}
\phi_i(x, b)&=&\int d^2{\vec k_{\bot}}~e^{-i\vec k_{\bot}
             \cdot \vec b}\phi_i(x, k_{\bot})\nonumber\\
     &=&\phi_i(x) \exp(-\frac{b^2}{4\beta^2})
\end{eqnarray}
where the parameter $\beta$ is taken as $\beta^2=4.0\;\mbox{GeV}^{-2}$ according to the discussion of Refs. \cite{jk1993, wy2002}.

The distribution amplitude $\phi_\pi(x)$ is for the twist-2 wave function, and $\phi^\pi_P(x)$ and $\phi^\pi_\sigma(x)$ are for twist-3 wave function, which are given by \cite{bra1990,bal1999,bbl2006}
\begin{eqnarray}
 \phi_\pi(x) &=&  6x(1-x) \left[1+a_2^\pi C_2^{3/2} (t) \right]\label{piw1}\\
 \phi_\pi^P(x) &=&  1+a_{2P} C_2^{1/2} (t) +a_{4P} C_4^{1/2} (t)  \\
 \phi_\pi^\sigma(x) &=& 6x(1-x)\left[1+a_{2\sigma}6(5x^2-5x+1)\right]\\
 \phi_\pi^t(x) &=&  6(1-2x)  \left[ 1+a_{2\sigma} 6 (10x^2-10x+1)  \right]     \label{piw}
 \end{eqnarray}
where $t=1-2x$.  The function $C$'s are Gegenbauer polynomials, which are defined by
 \begin{equation}
 \begin{array}{ll}
 C_2^{1/2} (t) = \frac{1}{2} (3t^2-1), & C_4^{1/2} (t) = \frac{1}{8}
 (35t^4-30t^2+3),\\
 C_2^{3/2} (t) = \frac{3}{2} (5t^2-1), & C_4^{3/2} (t) = \frac{15}{8}
 (21t^4-14t^2+1)
 \end{array}
 \end{equation}
 and the coefficients are \cite{bbl2006}
 \begin{eqnarray}
 &&a_2^\pi(\mu=1\mbox{GeV})=0.25\pm 0.15,  \nonumber\\
 && a_{2P}(\mu=1\mbox{GeV})=0.58\pm 0.23, \nonumber\\
 &&a_{4P}(\mu=1\mbox{GeV})=0.067\pm 0.055, \nonumber\\
 && a_{2\sigma}(\mu=1\mbox{GeV})=0.11\pm 0.04  \label{a-value}
 \end{eqnarray}
which correspond to $\mu_{\pi}=1.75$ GeV. The hard scale of $B$ meson decays is mainly around $\sqrt{m_B \Lambda_{\mbox{QCD}}}=1.1\sim 1.4$ GeV, which coincides with the scale of the parameters given in Eq. (\ref{a-value}), so the evolution effect of pion distribution amplitudes can be neglected.


\begin{thebibliography}{99}
\bibitem{QCDf1}M. Beneke, G. Buchalla, M. Neubert, and C.T. Sachrajda,``QCD Factorization for
   $B\to \pi\pi$ Decays: Strong Phases and CP Violation in the Heavy Quark Limit",
   Phys. Rev. Lett. 83, 1914 (1999).
\bibitem{QCDf2}M. Beneke, G. Buchalla, M. Neubert, and C.T. Sachrajda, ``QCD factorization for
   exclusive non-leptonic B-meson decays: general arguments and the case of heavy¨Clight final
   states", Nucl. Phys. B 591, 313 (2000).
\bibitem{QCDf3}M. Beneke, G. Buchalla, M. Neubert, and C.T. Sachrajda, `` QCD factorization in
   $B \to \pi K$, $\pi \pi$ decays and extraction of Wolfenstein parameters", Nucl. Phys. B 606, 245 (2001).
\bibitem{QCDf4}M. Beneke, M. Neubert, ``QCD factorization for $B \to P P$ and $B \to P V$ decays",
   Nucl. Phys. B 675, 333 (2003).
\bibitem{PQCD1}H. N. Li and H. L. Yu, ``Extraction of $V_{ub}$ from the Deacy $B\to \pi l\nu$",
         Phys. Rev. Lett. 74, 4388 (1995).
\bibitem{PQCD2}H. N. Li and H. L. Yu, ``PQCD analysis of exclusive charmless B meson decay spectra",
         Phys. Lett. B 353, 301 (1995).
\bibitem{PQCD3}H. N. Li, ``Applicability of perturbative QCD to $B \to D$ decays", Phys. Rev. D 52, 3958 (1995).
\bibitem{PQCD4}Y. Y. Keum, H. N. Li, and A. I. Sanda, ``Fat penguins and imaginary penguins in perturbative QCD",
         Phys. Lett. B 504, 6 (2001).
\bibitem{PQCD5}Y. Y. Keum, H. N. Li, and A. I. Sanda, ``Penguin enhancement and $B \to K \pi$ decays
         in perturbative QCD", Phys. Rev. D 63, 054008 (2001).
\bibitem{PQCD6}C. D. L\"{u}, K. Ukai, and M. Z. Yang, ¡®¡¯Branchingratio and $CP$ violation of $B\to\pi\pi$ decays
     in the perturbative QCD approach", Phys. Rev. D 63, 074009 (2001).
\bibitem{PQCD7}C. D. L\"{u}, M. Z. Yang, ``$B\to\pi\rho$, $\pi\omega$ decays in perturbative QCD approach",
    Eur. Phys. J. C 23, 275 (2002).
\bibitem{PW1}H.D. Politzer and M.B.Wise, ``Leading Logarithms of the Heavy Qaurk Masses In Processes With Light and
    Heavy Quarks", Phys. Lett. B 206, 681 (1988).
\bibitem{PW2}H.D. Politzer and M.B.Wise, ``Effective Field Theory Approach To Processes Involving Both Light and Heavy
    Fields",  Phys. Lett. B 208, 504 (1988).
\bibitem{GN}A.G. Grozin, M. Neubert, ``Asymptotics of heavy-meson form factors", Phys. Rev. D 55, 272 (1997).
\bibitem{LN}B.O. Lange and M. Neubert, ``Renormalization-Group Evolution of the $B$-Meson Light-Cone Distribution
   Amplitude", Phys. Rev. Lett. 91, 102001 (2003).
\bibitem{BL} G. P. Lepage and S. J. Brodsky, ``Exclusive processes in perturbative chromodynamics", Phys. Rev. D 22, 2157
(1980).
\bibitem{Yang2012}M.Z. Yang, ``Wave functions and decay constants of B and D mesons in the relativistic potential model",
    Eur. Phys. J. C 72, 1880 (2012).
\bibitem{LY2014}J.B. Liu and M.Z. Yang, ``Spectrum of the charmed and b-flavored mesons in the relativistic potential model", JHEP 07, 106 (2014).
\bibitem{LY2015}J.B. Liu and M.Z. Yang, ``Spectrum of higher excitations of B and D mesons in the relativistic potential model", Phys. Rev. D 91, 094004 (2015)
\bibitem{SY2017}H.K. Sun and M.Z. Yang, ``Decay constants and diftribution amplitudes of $B$ meson in the relatisvistic
    potential model", Phys. Rev. D 95, 113001 (2017).
\bibitem{SY2019}H.K. Sun and M.Z. Yang, ``Wave functions and leptonic decays of bottom mesons in the relativistic potential model", Phys. Rev. D 99, 093002 (2019).
\bibitem{liyu1996-1}H. N. Li and H.L. Yu, ``Perturbative QCD analysis of B meson decays", Phys. Rev. D 53, 2480 (1996).
\bibitem{luyang2003}C.D. L\"{u}, M.Z. Yang, ``B to light meson transition form factors calculated in perturbative QCD approach", Eur. Phys. J. C 28, 515 (2003).
\bibitem{liyu1996-2}H.N. Li and H.L. Yu, ``PQCD analysis of inclusive semileptonic decays of B mesons", Phys. Rev. D 53, 4970 (1996).
\bibitem{lihn2002}H.N. Li, ``Threshold resummation for exclusive B meson decays", Phys. Rev. D 66, 094010 (2002).
\bibitem{bra1990}V.M. Braun, I.E. Filyanov, ``Conformal Invariance and Pion Wave Functions of Nonleading Twist", Z Phys. C 48, 239 (1990).
\bibitem{bal1999}P. Ball, ``Theoretical update of pseudoscalar meson distribution amplitudes of higher twist: the nonsinglet case", J. High Energy Phys. 01, 010 (1999)
\bibitem{bf2001}M. Beneke, T. Feldmann, ``Symmetry-breaking corrections to heavy-to-light B meson form factors at large recoil", Nucl. Phys. B 592, 3¨C34 (2001).
\bibitem{wy2002}Z.T. Wei, M.Z. Yang, ``The systematic study of $B\to\pi$ form factors in PQCD approach and its reliability", Nucl. Phys. B 642, 263 (2002).
\bibitem{lsw2012}H.N. Li, Y.L. Shen, and Y.M. Wang, ``Next-to-leading-order corrections to $B\to\pi$ form factors in $k_T$ factorization", Phys. Rev. D 85, 074004 (2012).\bibitem{jk1993}R. Jakob, P. Kroll, ``The Pion form-factor: Sudakov suppressions and intrinsic transverse momentum",Phys. Lett. B 315, 463¨C470 (1993).
\bibitem{bbl2006}P. Ball, V.M. Braun and A. Lenz, ``Higher-twist distribution amplitudes of the K meson
    in QCD", J. High Energy Phys. 05, 004 (2006).
\bibitem{BF2004}M. Beneke, Th. Feldmann, ``Factorization of heavy-to-light form factors in soft-collinear effective theory", Nucl. Phys. B 685 (2004) 249.
\bibitem{barbar2011}P.A. Sanchez et. al (\rm{BarBar} Collaboration), ``Study of $B\to \pi l\nu$ and  $B\to \rho l\nu$ decays and determination of $|V_{ub}|$", Phys. Rev. D 83, 032007 (2011).
\bibitem{belle2011}H. Ha et al. (Belle collaboration), `` Measurement of the decay $B^0\to\pi^-\ell^+\nu$ ad determination of $|V_{ub}|$", Phys. Rev. D 83, 071101(R) (2011).
\bibitem{barbar2012}J.P. Lees et. al (\rm{BarBar} Collaboration), ``Branching fraction and form-factor shape measurements of exclusive charmless semileptonic $B$ decays, and determination of $|V_{ub}|$", Phys. Rev. D 86, 092004 (2012).
\bibitem{belle2013}A. Sibidannov et. al (Belle collaboration), ``Study of exclusive $B\to X_u\ell\nu$ decays and extraction of $|V_{ub}|$ using full reconstruction tagging at the Belle experiment", Phys. Rev. D 88, 032005 (2013).
\bibitem{lcsr}A. Bharucha, ``Two-loop corrections to the $B\to\pi$ form factor from QCD sum rules on the light-cone and $|V_{ub}|$", J. High Energy Phys. 05, 092 (2012).
\end{thebibliography}
\end{document}